\documentclass[twocolumn]{aastex631}
\shorttitle{T CrB Orbital Period Changes}
\shortauthors{Schaefer}
\graphicspath{{./}{figures/}}
\begin{document}
\title{Orbital Period Changes in Recurrent Nova T Corona Borealis Prove That It Is Not a Type {\rm I}a Supernovae Progenitor}

\author[0000-0002-2659-8763]{Bradley E. Schaefer}
\affiliation{Department of Physics and Astronomy,
Louisiana State University,
Baton Rouge, LA 70803, USA}

\begin{abstract}

T Corona Borealis (T CrB) is a recurrent nova and a symbiotic star that is commonly highlighted as the best case for being a progenitor of a Type Ia supernova (SNIa) within the framework of single-degenerate models.  This exemplar can be tested by measuring whether the white dwarf (WD) mass ($M_{\rm WD}$) is increasing over each eruption cycle.  This is a balance between the mass ejected during each nova event ($M_{\rm ejecta}$) and the mass accreted onto the WD between the nova events ($M_{\rm accreted}$).  I have used all 206 radial velocities from 1946--2024 to measure the orbital period just after the 1946 eruption to be $P_{\rm post}$=227.6043 days, while the steady orbital period change ($\dot{P}$) is ($-$3.1$\pm$1.6)$\times$10$^{-6}$.  I have used my full 213,730 magnitude $B$ and $V$ light curve from 1842--2025 to measure the times of maximum brightness in the ellipsoidal modulations to construct the $O-C$ from 1866--1946.  I fit the broken parabola shape, to find the orbital period immediately before the 1946 eruption to be $P_{\rm pre}$=227.4586 days.  The orbital period changed by $\Delta P$=$+$0.146$\pm$0.019 days.  With Kepler's Law, conservation of angular momentum, and the well-measured binary properties, the ejecta mass in 1946 is 0.00074$\pm$0.00009 M$_{\odot}$.  $M_{\rm accreted}$ is reliably measured to be 1.38$\times$10$^{-6}$ M$_{\odot}$ from the accretion luminosity.  $M_{\rm ejecta}$ is larger than $M_{\rm accreted}$ by 540$\times$, so $M_{\rm WD}$ is {\it decreasing} every eruption cycle.  T CrB can never become a SNIa.

\end{abstract}

\section{INTRODUCTION}

The Type Ia supernova (SNIa) progenitor problem is the question of the nature of the system before the explosion, with this being one of the grand-challenge problems of broad importance through many areas of astrophysics.  The progenitor problem is to determine the nature of the companion star, and the detailed explosion mechanism in the white dwarf.  Since the 1980s, vigorous debate has been on-going, with many competing model scenarios, plus a wide array of observational evidences (for reviews, see Livio 2000, Maoz, Mannucci, \& Nelemans 2014, Livio \& Mazzali 2018, Patat \& Hallakoun 2018, and Ruiter \& Seitenzahl 2025).  The progenitor certainly is a binary star for which one component must be a white dwarf (WD) with carbon/oxygen (CO) composition that somehow increases its mass until a thermonuclear runaway explosion is triggered.  (If the WD has an oxygen/neon (ONe) composition, then there is not enough nuclear energy available to power the observed SNIa events.)  Progenitor scenarios are popularly divided into two classes, denoted by the number of WDs (degenerate stars) in the binary.  The single-degenerate (SD) models have the binary consisting of {\it one} WD plus a relatively normal star in close orbit feeding gas onto the WD by accretion until the WD explodes.  The double-degenerate (DD) models have the binary consisting of {\it two} WDs in close orbit, in-spiraling by gravitational radiation until the two WDs collide/merge so as to then explode.

For SD models, the most popular have always been that the progenitor is either or both of a recurrent nova (RN) or a symbiotic star (SS).  RNe are classical nova binaries that have a recurrence timescale of under one century, for which only eleven are known in our Milky Way (Schaefer 2010, Schaefer et al. 2022).  To achieve such a fast recurrence timescale, the RNe must have near-maximal WD mass ($M_{\rm WD}$) and a near-maximal accretion rate ($\dot{M}$), and this is exactly the state that naively must result in the WD reaching the Chandrasekhar mass and exploding as a SNIa.  SSs are binaries where one star is a WD accreting gas from a red giant companion star that has a massive stellar wind (Miko{\l}ajewska 2010, Munari 2019, Miko{\l}ajewska et al. 2021).  This accretion could be either by Roche lobe overflow or from the stellar wind.  Most SSs have low mass WDs ($M_{\rm WD}$$\sim$0.6 M$_{\odot}$), so models must make the difficult argument that the WD mass is increasing.  Nevertheless, four sister binaries are both symbiotic stars and recurrent novae, labeled as Symbiotic Recurrent Novae (SyRNe), including T CrB, RS Oph, V3870 Sgr, and V745 Sco.

T CrB is the premier exemplar of the SD scenario (e.g., Kenyon et al. 1993, Hachisu, Kato, \& Nomoto 1999).  As an RN, it had observed classical nova eruptions in the years 1217, 1787, 1866, 1946, plus one that is likely to erupt any month now. (Schaefer 2023a, 2023b).  As an SS, the companion star is an M4.1 III red giant in a 227.58 day orbital period ($P$).  The properties of T CrB are now well-measured, as summarized in Schaefer (2010), Schaefer (2023a), Planquart, Jorissen, \& Van Winckel (2025), and Hinkle et al. (2025).  $M_{\rm WD}$ is measured from the radial velocity curve to be variously 1.37$\pm$0.01, 1.31$\pm$0.05, and 1.32$\pm$0.10 M$_{\odot}$, depending mainly on whether the {\it Gaia} distance is adopted.  The red giant companion has a mass of $M_{\rm comp}$ variously given as 0.69$^{+0.02}_{-0.01}$, 0.419$^{+0.002}_{-0.004}$, and 0.98$\pm$0.31 M$_{\odot}$.  For much of the $\sim$80 year interval between eruptions, T CrB is in a low state of accretion that averages out to $\dot{M}_{\rm low}$=3.2$\times$10$^{-9}$ M$_{\odot}$ yr$^{-1}$.  In the decade before each nova eruption and the decade after each nova eruption, T CrB is in a high state\footnote{This high state is labeled in some papers as a `super-active phase'.  I think that this is a poor name, because the word `super' conjures up extreme cases with extraordinarily high $\dot{M}$ and it conjures up `super-novae'.  Rather, `super' is inappropriate for a case where the system only brightens slowly by 1 mag.} with the accretion rate equal to $\dot{M}_{\rm high}$=6.4$\times$10$^{-8}$ M$_{\odot}$ yr$^{-1}$.  T CrB is the most famous RN, and is the most famous SS, with the upcoming eruption only increasing the importance of the system.  For the progenitor problem, T CrB is the keystone of SD models.

How can we test whether T CrB itself will evolve until its WD explodes as a SNIa?  One way to test this is to see whether the underlying WD composition is CO or ONe.  If T CrB is a `neon nova', where the nova ejecta has a high over-abundance of neon (with respective to solar abundance), then the bulk neon can only come from dredged-up neon-rich mass on the surface of the underlying ONe WD.  Any dredged-up neon can only be detected by the neon emission lines visible only during the nebular phase late in the nova light curve.  Soon after the 1946 eruption, a strong [Ne III] emission line at 3869~\AA~ was seen brighter than the [O III] 5007~\AA~line (Bloch \& Tcheng 1953).  The observed line flux ratio for T CrB is $F_{3869}/F_{5007}$=1.4.  These flux ratios are strongly correlated with the neon abundances relative to solar, for 37 galactic novae, as measured from full radiative transfer abundance analyses.  {\it All} neon novae (as explicitly identified from independent abundance analyses) have observed extinction-corrected ratios $\ge$0.3, with all of these neon novae having neon abundances $\ge$10$\times$ solar.  {\it All} novae that are not recognized neon-novae have the ratio $\le$0.25, while having neon abundances of $\le$7$\times$ solar.  So the neon-to-oxygen line ratio appears to be an excellent and reliable empirical criterion for a neon nova, and hence for an ONe WD.  T CrB has its observed $F_{3869}/F_{5007}$ close to the median value for all recognized neon novae.  Planquart et al. (2025) prove that these neon and oxygen emission lines in T CrB both come only from the same region, far out in the expanding shell.  With these observations, T CrB is apparently a neon nova, has an ONe WD, and cannot become a SNIa.  Before this conclusion can be finalized, there must be a full radiative transfer analysis of many emission lines as observed late in the tail of the upcoming nova eruption.

Another way to test T CrB as a SD progenitor is to measure whether $M_{\rm WD}$ is increasing or decreasing over each eruption cycle.  The long-term rise or fall of the WD mass is a balance between the mass ejected during each eruption ($M_{\rm ejecta}$) and the mass gained by accretion between eruptions ($M_{\rm accreted}$).  In general and for T CrB in particular, we have good estimates of $M_{\rm accreted}$.  So the test comes down to measuring $M_{\rm ejecta}$ and seeing whether $M_{\rm ejecta}$$>$$M_{\rm accreted}$?  

For measuring $M_{\rm ejecta}$, all the traditional methods (usually involving the measure of the luminosity of one of the hydrogen Balmer lines soon after peak) have horrible problems arising from big uncertainties in the distance, extinction, shell volume, filling factors, opacity in the lines, gas density, and gas temperatures, plus accounting for the wide range of conditions simultaneously present in the shell.  In practice, Schaefer (2011) identifies six specific sources of errors, each with from 1-to-3 orders-of-magnitude uncertainty, for a combined error of around 3 orders-of-magnitude.  This is illustrated by the four published ejecta masses for U Sco varying over a range of 720$\times$ (Schaefer 2011, Appendix A), and the four published masses for V445 Pup varying by 7 orders-of-magnitude (Schaefer 2025).  And published theory has similar large disagreements over the order-of-magnitude expected for $M_{\rm ejecta}$ (Schaefer 2011, Appendix A).  So the reality is that none of the many published measures of $M_{\rm ejecta}$ are valid to within even orders-of-magnitude.  

There is one exception to this, and this accurate and reliable method is only applicable to a small handful of novae.  This good method is to measure the sudden change of the orbital period $P$ across the nova eruption as $\Delta P$.  The ejecta mass is directly given from Kepler's Law and the conservation of angular momentum.  The measure of $M_{\rm ejecta}$ comes down to a reliable timing experiment, completely independent of distances, extinctions, filling factors, and temperatures.  There are two problems with this reliable method.  The first problem is that it formally only returns a lower limit on $M_{\rm ejecta}$, because the `frictional angular momentum loss' (FAML) as the companion star plows through the nova ejecta is poorly known.  Fortunately, for the case of of RNe, with necessarily small $M_{\rm ejecta}$ and high ejecta velocities, the FAML must be negligibly small, so the lower limit becomes an equality.  The second problem is that this method can be applied to few novae, because few have any accurate pre-eruption period ($P_{\rm pre}$) that can be measured from archival data.  Without $P_{\rm pre}$ and the post-eruption period ($P_{\rm post}$), we cannot get a $\Delta P$=$P_{\rm post}$$-$$P_{\rm pre}$ measure.  Further, the $\Delta P$ method is only useful for a nova that happens to have the observed $\Delta P$ value substantially positive.  Recently, for the unique case of the helium nova V445 Pup with a giant companion star, I have measured the $\Delta P$/P to be $+$935$\pm$27 ppm, so the $M_{\rm ejecta}$ must be $\gg$0.001 M$_{\odot}$ (Schaefer 2025).  With vast effort as a career-long program started in 1981, I have measured $\Delta P$ across 14 nova eruptions on 11 nova systems (Schaefer 2023c, Schaefer 2025).  I now have measures for just 5 novae (including T CrB) that are useful for constraining $M_{\rm ejecta}$, and these are likely to be all that are possible for long into the future.

The plan for this paper starts with collect all existing radial velocity data from 1946--2025 and all photometric data 1866--2025 so as to measure the orbital period over the the entire last two eruption cycles.  From this, I will derive the sudden period change across the 1946 nova event ($\Delta P$), and the slow and steady period change between eruptions ($\dot{P}$).  A convenient and insightful depiction of these measured and modeled period changes is with the traditional $O-C$ diagram, plotting the deviations of some orbital conjunction markers (as compared to some fiducial constant period) over the years.  An unchanging $P$ would appear as a straight line, a steady $\dot{P}$ would appear as a parabola, and a sudden $\Delta P$ would appear as a sharp kink in the slope.  With the measured $\Delta P$ and the binary properties, I will derive the $M_{\rm ejecta}$ from the 1946 nova.  This will then be compared to $M_{\rm accreted}$, with the conclusion as to whether the T CrB $M_{\rm WD}$ is gaining or losing mass over the entire eruption cycle.

\section{T C\lowercase{r}B PERIOD CHANGES}

Two methods can be used to measure the orbital phase markers for use in measuring $P$ and its changes.  The first method uses the radial velocity of the red giant spectral bands to mark the instant of conjunction as when the velocity is zero (i.e., at the $\gamma$-velocity).  This method is relatively accurate, and we have good measures from 1946--2025.  The second method is from the usual ellipsoidal modulations seen for the red giant at half the orbital period (i.e., the ellipsoidal modulations appear as a sinewave in the B and V magnitudes with a period near to 113.79 days), with the conjunctions marked by the mimima in the light curve.  This method is poorer in accuracy than the radial velocity method, but it has the only information about the period from 1866--1946, as required to get $P_{\rm pre}$.

\subsection{Period Changes from the Radial Velocity Curve}

Good radial velocity measures have been reported by Sanford (1947), Kraft (1958), Kenyon \& Garcia (1986), Fekel et al. (2000), Planquart et al. (2025) and Hinkle et al. (2025).  All of these produce a perfect sinewave radial velocity curve, consistent amongst all observers.  No one has analyzed any data set to look for a steady period change with $\dot{P}$.

I have taken {\it all} radial velocity measures and fitted them to a sinewave where the period is allowed to steadily change.  For this, I have used the compilation in Hinkle et al. (2025), adopting their systemic offsets for each observer.  (The last three measures were not included as Hinkle quotes a zero weight for these points.)  The one-sigma error bars were taken as the RMS of the residuals from the best-fit for each observer.  The model fit consisted of all 206 radial velocities being compared to a sinewave with a constant $\dot{P}$ in the usual chi-square manner, where the important fit parameters were the epoch (in heliocentric Julian days, HJD) for the maximum radial velocity, the period at that epoch, and the $\dot{P}$ (taken as a dimensionless quantity that can be expressed as s/s).  The zero phase corresponds to one of the elongations of the orbit, while the conjunctions are marked by the times when the velocity passes through the middle $\gamma$-velocity.  Incidental fit parameters include the K-velocity (fitted to 24.04 km s$^{-1}$) and the $\gamma$-velocity (fitted to $-$27.94 km s$^{-1}$).  The best fit values were taken from the point with the minimum chi-square, while the 1-$\sigma$ confidence error bars were taken as the region of parameter space within which the chi-square is 1.00 larger than that of the minimum.  

For the model, I took the fiducial epoch to be near the first measure of Planquart et al. (2025) in 2011, because it was near the weighted center of time for all the observations.  With this, the epoch of the fitted maximum radial velocity is at HJD 2455427.82$\pm$0.13.  At that time, the fitted period is 227.5320 days.

My best fitting model has a moderately significant negative $\dot{P}$, with the period {\it decreasing} over time.  My best fitting value for $\dot{P}$ is $-$3.1$\times$10$^{-6}$ in dimensionless units.  The 1-$\sigma$ range for $\dot{P}$ is from $-$1.5$\times$10$^{-6}$ to $-$4.7$\times$10$^{-6}$.  The possibility of $\dot{P}$ equaling zero is rejected at the 2-$\sigma$ confidence level.

With this model fit, the orbital period in 1946 was 227.6043 days, and in 2025 was 227.5147 days.  With this, $P$ decreased by 0.0896 days in 127 orbits, or 1.02 minutes per orbit.  The value of $\log(|\dot{P}|)$ is -5.5, with the magnitude of $|\dot{P}|$ being two orders-of-magnitude larger than for all other known nova binaries (Schaefer 2023c).  This huge negative number can only arise from a massive loss of angular momentum in the binary.  Currently, there is no understanding or physical model that can explain such a huge angular momentum loss (AML).  Presumably, the mechanism must involve the red giant stellar wind carrying away angular momentum from the binary.

The observed $\dot{P}$ is just the sum of the effect caused by the mass transfer within the binary, $\dot{P}_{\rm mt}$, plus the effect caused by some unknown mechanism for steady loss of angular momentum by the binary, $\dot{P}_{\rm AML}$, with 
\begin{equation}
\dot{P} = \dot{P}_{\rm mt} + \dot{P}_{\rm AML}.
\end{equation}
For essentially all cataclysmic variable, with $M_{\rm WD}$$>$$M_{\rm comp}$, $\dot{P}_{\rm mt}$ must be positive.  The angular momentum of the binary can only be lost, as there is nothing outside the binary that can provide a torque, so necessarily $\dot{P}_{\rm AML}$ will be negative.  The resultant value of $\dot{P}$ will depend on the balance between mass transfer and AML.  At this time, the AML cannot be measured or predicted, but we at least know that $\dot{P}_{\rm AML}$$\le$0.  With this, we know that 
\begin{equation}
\dot{P} \le \dot{P}_{\rm mt}.
\end{equation}
For conservative mass transfer, the effects of the mass transfer are
\begin{equation}
\dot{P}_{\rm mt} = (3 P \dot{M}/M_{\rm comp})(1-q),
\end{equation}
where $q$ is the usual mass ratio $M_{\rm comp}$/$M_{\rm WD}$.  With $q$ always $<$1 for novae with Roche lobe overflow, $\dot{P}_{\rm mt}$ will always be positive.

For T CrB, adopting $M_{\rm WD}$=1.32 M$_{\odot}$, $M_{\rm comp}$=0.98 M$_{\odot}$, and $\dot{M}_{\rm high}$=6.4$\times$10$^{-8}$ M$_{\odot}$ yr$^{-1}$, then $\dot{P}_{\rm mt}$ is $+$3.1$\times$10$^{-8}$ in dimensionless units.  Then with Equation 2, we know that the observed $\dot{P}$ must always be $<$$+$3.1$\times$10$^{-8}$.  This limit will be critical below for evaluating $\Delta P$.

What is $\dot{P}$ from 1866--1946?  The $O-C$ curve for the ellipsoidal modulation are too poor to usefully constrain $\dot{P}$.  The $\dot{P}$ values change sharply across a nova eruption for both U Sco and T Pyx.  Nevertheless, the best estimate for before the 1946 eruption is $-$3.1$\times$10$^{-6}$, with a range of $-$1.5$\times$10$^{-6}$ to $-$4.7$\times$10$^{-6}$.  In all cases, we have the strong limit that $\dot{P}$$<$$+$3.1$\times$10$^{-8}$.

\subsection{Period Changes from the Ellipsoidal Modulations}

The position of the stars in their orbit can also be demarcated by the brightness of the red giant star.  The red giant star is somewhat stretched out of round, into `pointy' shape at the inner Lagrange point where the gas falls off the companion.  When viewed from the `side', as when the companion is at the elongations of its orbit as viewed from Earth, we see the large area of the broadside, so the star appears relatively bright.  When viewed with the pointy side aimed near to Earth, as when the companion is at either conjunction in its orbit as viewed from Earth, we see the side with a relatively small cross-sectional, so the star appears relatively faint.  As the out-of-round red giant orbits the WD, we will alternatively see the broadside or the narrow side, making the star's brightness oscillate up and down, roughly as a sinewave in the light curve, with a periodicity of half the orbital period.  This well known type of variability in binary stars is called the ellipsoidal modulation.

\begin{figure*}
	\includegraphics[width=2.1\columnwidth]{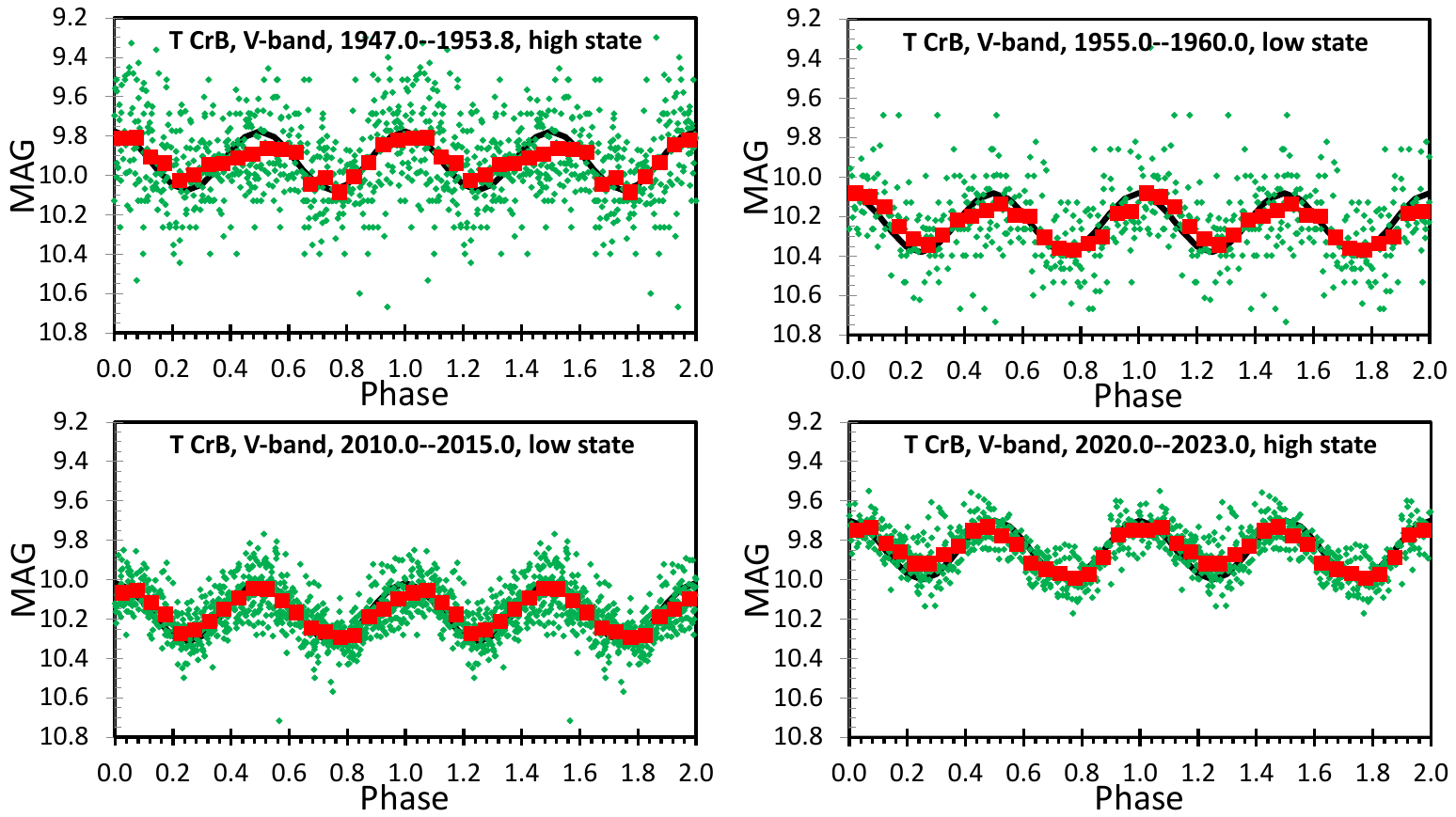}
    \caption{Ellipsoidal modulations in $V$-band after the 1946 eruption.  These light curves are folded with period 227.5687 days and an epoch of the best fit $T_{\rm max}$.  Phases 0.0, 1.0, and 2.0 correspond to the times of elongation in the orbit, when the red giant is maximally to one side of the white dwarf, and when the radial velocity is at its maximum value.  The zero phase is the time of peak brightness in the light curve (with the smallest magnitude).  Phases 0.5 and 1.5 are for the times of elongation with the red giant maximally on the other side of the WD.  Phases 0.25 and 1.25 are for the times of superior conjunctions in the orbit, with the red giant {\it behind} the WD, when the radial velocity is decreasing through the $\gamma$-velocity.  Phases 0.75 and 1.75 are for the times of inferior conjunctions in the orbit, with the red giant {\it in front of} the WD, when the radial velocity is increasing through the $\gamma$-velocity.  The $V$-band magnitudes are shown as green dots, many of which are averages over 0.01 year bins, representing 2 to 200 input magnitudes.  These individual points are phase averaged into bins each 0.05 wide in phase, as shown by the red squares.  My full light curve with 213,730 fully-modern Johnson $B$ and $V$ magnitudes from 1842--2022 are explicitly listed in Schaefer (2023a) and this is supplemented by recent data collected by the AAVSO.  The sinewave fit is shown by the black curve.  A large amount of the scatter is due to the usual `flickering' intrinsic to T CrB, whose variability has large power on all timescales (see Figure 13 of Schaefer 2023a).  With a chi-square fit to a sinewave, the 1-$\sigma$ uncertainty in the derived $T_{\rm max}$ is $\pm$1.0, $\pm$2.2, $\pm$0.8, and $\pm$0.8 days for the four intervals in time order.  A major point in showing these folded light curves is so that the ellipsoidal modulation can be seen in detail.  Another major point is to illustrate the sinewave fits from which the $T_{\rm max}$ are measured for the $O-C$ curve.} 
\end{figure*}

\begin{figure*}
	\includegraphics[width=2.1\columnwidth]{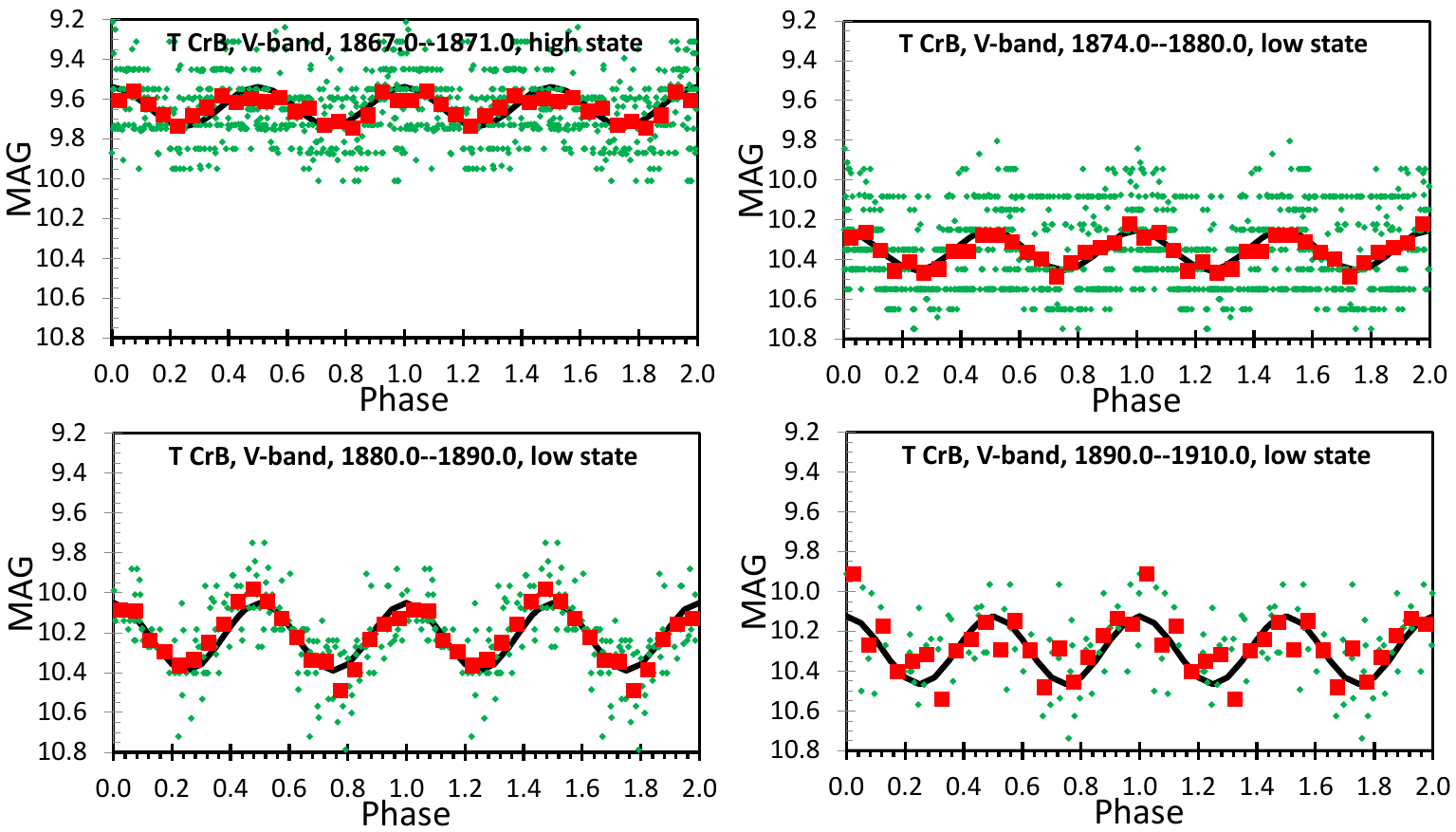}
    \caption{Ellipsoidal modulations in $V$-band before the 1946 eruption.  The details of this figure are the same as for Figure 1.  The interval 1890--1910 has relatively few points, so the phase binned points (in red squares) have relatively large scatter, and $T_{\rm max}$ has a formal uncertainty of $\pm$5.5 days.  For the earliest two intervals, the formal uncertainty in $T_{\rm max}$ is $\pm$2.8 and $\pm$2.1 days.  This is important because those two points have the largest lever arm for proving that the $O-C$ curve had a large upward kink in 1946.  Critically, these first two points are around 20 days {\it late} compared to that required for the $\Delta P$=0 case.  } 
\end{figure*}

For T CrB, the ellipsoidal modulation is roughly 0.3 mag in full amplitude, with a sinewave shape for the light curve.  Many examples of the light curve, in $V$, are shown in Figures 1 and 2.  These light curves are folded on the orbital period, taken with $P$=227.5687 days and with epoch of the time of peak brightness, $T_{\rm max}$, as fitted for each time interval.  The maxima in the light curves are near the times of orbital elongation, i.e., when the red giant is viewed broadside.  The minima in the light curves are near the times of conjunction.   Each particular $T_{\rm max}$ corresponds to the one time of maximum light when the radial velocity is also at its maximum.

This modulation is exactly tied to the orbital period, because the pointy shape of the companion must always be at the inner Lagrangian point.  With this, the phase of the ellipsoidal effect can be used to measure the times of conjunction and hence the period.  The orbital phase can be measured by fitting a sinewave to the light curve and taking the time of maximum brightness.  For a given time interval within the full light curve, I have made chi-square fits to a sinewave for the $B$ or $V$ light curve magnitudes.  The times of maximum light ($T_{\rm max}$), selected for a peak in the middle of the interval, are listed in Table 1.  The table also specifies the year range and the band for each fit.  The state of the interval is identified in the third column.  My fitted $T_{\rm max}$ with error bars are in the fifth column, while the corresponding calendar year is back in the second column.  I have $V$-band data from 1842 to last month, plus $B$-band data from 1890 to last month, but I have only fitted sinewaves for intervals for which there are enough magnitudes to provide a useful accuracy.  I have also avoided time intervals of transition between the states, and I have avoided intervals involving the primary and secondary nova eruptions.  I have 18 $V$-band times for intervals stretching from 1867 to 2025, plus 13 $B$-band times for intervals from 1900 to 2025.

\begin{table*}
	\centering
	\caption{$T_{\rm max}$ and $O-C$ from Ellipsoidal Modulation from 1867--2025}
	\begin{tabular}{lllllrcrcc}
		\hline
		   &     &      &    &      &  Observed     &     RV    &    Measured    &     Adopted     &    Final \\
		Year Range   &   Year   &   State   &  Band  &  $T_{\rm max}$ (HJD)    &  $O-C$ (d)     &     $O-C$ (d)    &    Offset (d)    &     Offset (d)     &    $O-C$ (d) \\
		\hline
2023.6	 --	2025.2	&	2024.34	&	Dip	&	V	&	2460435.6	$\pm$	1.1	&	0.70	&	-1.55	$\pm$	0.12	&	2.25	&	2.25	$\pm$	1.00	&	-1.55	$\pm$	1.49	\\
2020.0	 --	2023.2	&	2021.85	&	High	&	V	&	2459524.3	$\pm$	0.8	&	-0.32	&	-1.45	$\pm$	0.10	&	1.13	&	2.36	$\pm$	1.50	&	-2.68	$\pm$	1.70	\\
2016.0	 --	2020.0	&	2018.12	&	High	&	V	&	2458163.5	$\pm$	0.9	&	4.29	&	-1.06	$\pm$	0.09	&	5.35	&	2.36	$\pm$	1.50	&	1.93	$\pm$	1.75	\\
2010.0	 --	2015.0	&	2012.51	&	Low	&	V	&	2456113.7	$\pm$	0.8	&	2.61	&	-0.68	$\pm$	0.11	&	3.29	&	4.48	$\pm$	0.50	&	-1.87	$\pm$	0.94	\\
2000.0	 --	2010.0	&	2005.66	&	Low	&	V	&	2453611.6	$\pm$	0.7	&	3.76	&	-0.30	$\pm$	0.15	&	4.06	&	4.48	$\pm$	0.50	&	-0.71	$\pm$	0.86	\\
1990.0	 --	2000.0	&	1995.07	&	Low	&	V	&	2449742.8	$\pm$	0.8	&	3.63	&	0.13	$\pm$	0.13	&	3.50	&	4.48	$\pm$	0.50	&	-0.85	$\pm$	0.94	\\
1980.0	 --	1990.0	&	1985.10	&	Low	&	V	&	2446101.8	$\pm$	0.9	&	3.73	&	0.35	$\pm$	0.20	&	3.38	&	4.48	$\pm$	0.50	&	-0.75	$\pm$	1.03	\\
1970.0	 --	1980.0	&	1975.14	&	Low	&	V	&	2442464.3	$\pm$	1.2	&	7.33	&	0.38	$\pm$	0.21	&	6.95	&	4.48	$\pm$	0.50	&	2.85	$\pm$	1.30	\\
1960.0	 --	1970.0	&	1965.17	&	Low	&	V	&	2438821.2	$\pm$	1.1	&	5.33	&	0.23	$\pm$	0.29	&	5.10	&	4.48	$\pm$	0.50	&	0.85	$\pm$	1.21	\\
1955.0	 --	1960.0	&	1957.69	&	Low	&	V	&	2436090.1	$\pm$	2.2	&	5.05	&	0.00	$\pm$	0.39	&	5.05	&	4.48	$\pm$	0.50	&	0.58	$\pm$	2.26	\\
1947.0	 --	1953.8	&	1950.20	&	High	&	V	&	2433354.5	$\pm$	1.0	&	0.28	&	-0.33	$\pm$	0.55	&	0.61	&	2.36	$\pm$	1.50	&	-2.08	$\pm$	1.80	\\
1938.0	 --	1940.0	&	1938.37	&	High	&	V	&	2429032.9	$\pm$	3.3	&	2.48	&	...			&	...	&	2.36	$\pm$	1.50	&	0.12	$\pm$	3.63	\\
1925.0	 --	1927.0	&	1925.91	&	Low	&	V	&	2424481.7	$\pm$	4.7	&	2.66	&	...			&	...	&	4.48	$\pm$	0.50	&	-1.82	$\pm$	4.73	\\
1910.0	 --	1925.0	&	1914.69	&	Low	&	V	&	2420384.3	$\pm$	6.7	&	1.49	&	...			&	...	&	4.48	$\pm$	0.50	&	-2.98	$\pm$	6.72	\\
1890.0	 --	1910.0	&	1898.52	&	Low	&	V	&	2414480.6	$\pm$	5.5	&	14.58	&	...			&	...	&	4.48	$\pm$	0.50	&	10.10	$\pm$	5.52	\\
1880.0	 --	1890.0	&	1883.55	&	Low	&	V	&	2409010.7	$\pm$	3.4	&	6.33	&	...			&	...	&	4.48	$\pm$	0.50	&	1.85	$\pm$	3.44	\\
1874.0	 --	1880.0	&	1877.34	&	Low	&	V	&	2406742.8	$\pm$	2.1	&	14.11	&	...			&	...	&	4.48	$\pm$	0.50	&	9.64	$\pm$	2.16	\\
1867.0	 --	1871.0	&	1868.61	&	High	&	V	&	2403554.6	$\pm$	2.8	&	11.88	&	...			&	...	&	2.36	$\pm$	1.50	&	9.52	$\pm$	3.18	\\
2023.6	 --	2025.2	&	2024.34	&	Dip	&	B	&	2460434.5	$\pm$	2.6	&	-0.40	&	-1.55	$\pm$	0.12	&	1.15	&	1.15	$\pm$	1.00	&	-1.55	$\pm$	2.79	\\
2016.0	 --	2023.0	&	2019.98	&	High	&	B	&	2458840.5	$\pm$	2.1	&	-1.42	&	-1.20	$\pm$	0.09	&	-0.22	&	-0.22	$\pm$	1.00	&	-1.20	$\pm$	2.33	\\
2010.0	 --	2015.0	&	2012.51	&	Low	&	B	&	2456114.5	$\pm$	1.0	&	3.41	&	-0.68	$\pm$	0.11	&	4.09	&	3.75	$\pm$	0.96	&	-0.35	$\pm$	1.38	\\
2000.0	 --	2010.0	&	2008.14	&	Low	&	B	&	2454518.8	$\pm$	1.3	&	0.69	&	-0.43	$\pm$	0.13	&	1.12	&	3.75	$\pm$	0.96	&	-3.06	$\pm$	1.61	\\
1990.0	 --	2000.0	&	1995.07	&	Low	&	B	&	2449742.9	$\pm$	1.3	&	3.73	&	0.13	$\pm$	0.19	&	3.60	&	3.75	$\pm$	0.96	&	-0.02	$\pm$	1.61	\\
1980.0	 --	1990.0	&	1984.47	&	Low	&	B	&	2445873.7	$\pm$	2.6	&	3.20	&	0.35	$\pm$	0.20	&	2.85	&	3.75	$\pm$	0.96	&	-0.55	$\pm$	2.77	\\
1970.0	 --	1980.0	&	1973.90	&	Low	&	B	&	2442011.3	$\pm$	5.3	&	9.47	&	0.37	$\pm$	0.22	&	9.10	&	3.75	$\pm$	0.96	&	5.71	$\pm$	5.39	\\
1960.0	 --	1970.0	&	1963.91	&	Low	&	B	&	2438363.9	$\pm$	3.6	&	3.17	&	0.20	$\pm$	0.30	&	2.97	&	3.75	$\pm$	0.96	&	-0.59	$\pm$	3.73	\\
1955.0	 --	1960.0	&	1957.68	&	Low	&	B	&	2436087.6	$\pm$	4.1	&	2.55	&	0.00	$\pm$	0.39	&	2.55	&	3.75	$\pm$	0.96	&	-1.20	$\pm$	4.21	\\
1930.0	 --	1935.0	&	1932.76	&	Low	&	B	&	2426984.2	$\pm$	3.3	&	1.90	&	...			&	...	&	3.75	$\pm$	0.96	&	-1.85	$\pm$	3.44	\\
1920.0	 --	1930.0	&	1926.56	&	Low	&	B	&	2424720.7	$\pm$	6.5	&	14.09	&	...			&	...	&	3.75	$\pm$	0.96	&	10.33	$\pm$	6.57	\\
1910.0	 --	1920.0	&	1914.71	&	Low	&	B	&	2420392.8	$\pm$	2.9	&	9.99	&	...			&	...	&	3.75	$\pm$	0.96	&	6.24	$\pm$	3.05	\\
1900.0	 --	1910.0	&	1905.35	&	Low	&	B	&	2416973.8	$\pm$	2.6	&	4.52	&	...			&	...	&	3.75	$\pm$	0.96	&	0.77	$\pm$	2.77	\\
		\hline
	\end{tabular}	
	
\end{table*}

\begin{figure*}
	\includegraphics[width=2.1\columnwidth]{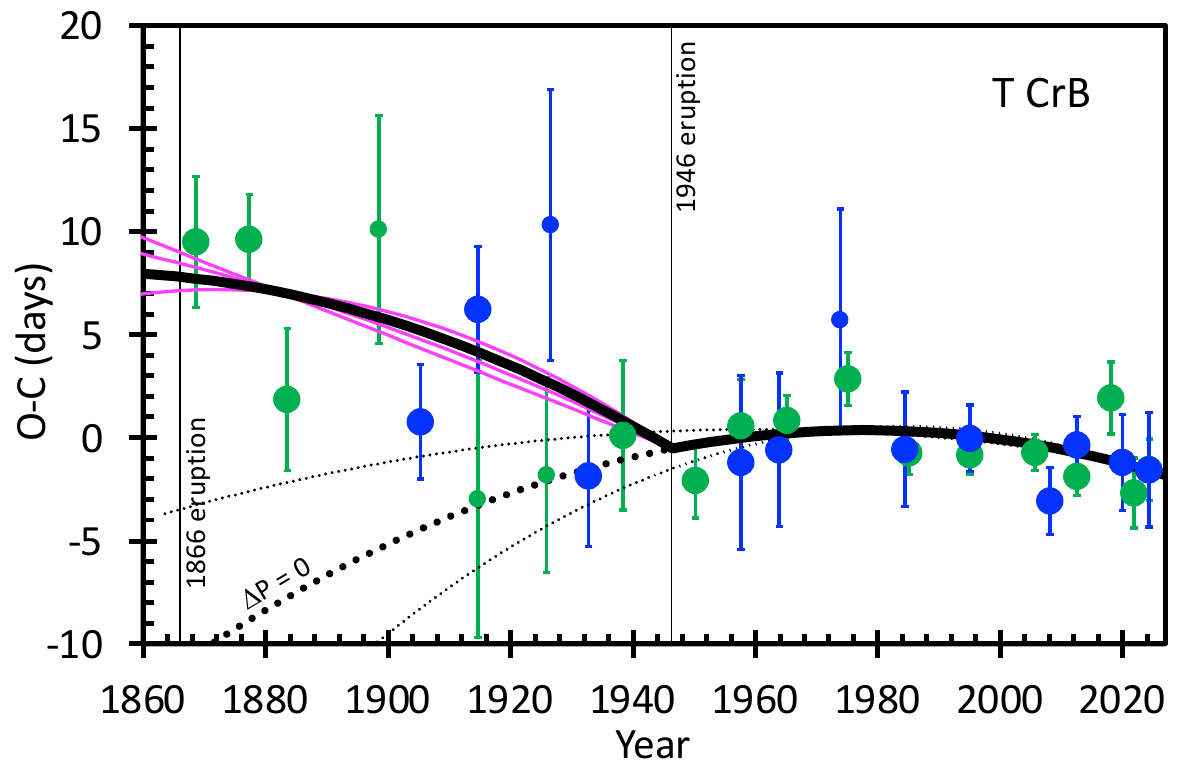}
    \caption{$O-C$ curve for T CrB.  The blue and green dots are the final $O-C$ values with one-sigma error bars for the $B$ and $V$ measures respectively (see Table 1), with smaller dots for measures with larger error bars.  The best-fitting curve for the radial velocity fit is the thick black curve from 1946--2025.  The dotted black curves show the extensions to earlier than 1946 for the $\Delta P$=0 case, with the thick dotted curve for the best fit $\dot{P}_{\rm post}$, and the flanking thin dotted curves representing the $\Delta P$=0 case with the 1-$\sigma$ range in the curvature.  All of the $O-C$ measures before 1910 are significantly and substantially above the $\Delta P$=0 case.  That is, T CrB certainly does not have a small or near-zero $\Delta P$.  The best fit model must be a broken parabola, where the steady period changes between eruptions make for parabolic segments with a sharp kink in 1946.  The parabola from 1866--1946 is accurately anchored in the year 1946 by the radial velocity data.  The only adjustable fit parameters for the 1866--1946 parabola are then the orbital period immediately before the 1946 eruption ($P_{\rm pre}$) and the curvature ($\dot{P}_{\rm pre}$).  With this, the best-fitting model with $\dot{P}_{\rm pre}$ equal to the best-fit $\dot{P}_{\rm post}$ is represented by the thick black curve from 1866 to 2025.  The flanking magenta curves represent the cases where the 1-$\sigma$ variations in the curvature are followed.  The nearly straight magenta curve represents the maximum possible curvature.  In all cases, there must be a sharp kink in 1946.  Visually, this can be seen because all the measures before 1910 are up to 20 days {\it late} as compared to any $\Delta P$=0 solution, and this is highly significant.  The kink is upward, so the sudden period change in 1946 must be positive, meaning that the period $\it increased$ due to the nova.  Quantitatively, the orbital period change is $+$0.146$\pm$0.019 days, and certainly $>$0.121 days.  This $\Delta P$ is huge, being two orders-of-magnitude larger than for any other measured nova.  This huge $\Delta P$ can only arise from a huge mass of ejecta. } 
\end{figure*}

My lists of $T_{\rm max}$ can be used to determine the $P$.  The best way to do this is with a traditional $O-C$ plot.  The $O-C$ value is $T_{\rm max}-T_{\rm model}$, where the model times are taken from the fiducial ephemeris of Fekel et al. (2000).  The error bars for $O-C$ are the same as the error bars for $T_{\rm max}$.  

For each of these intervals from 1946--2025, the model for the radial velocities produces a time of conjunction for the same conjunction as for $T_{\rm max}$.  Each time of zero phase from the radial velocity curve then makes for an $O-C$ value, as listed in the seventh column of Table 1.

In a comparison between the $T_{\rm max}$ as observed from the ellipsoidal effect and the times of zero phase as predicted from the radial velocity fits, the $T_{\rm max}$ values are systematically offset by up to 7 days.  That is, the ellipsoidal peaks come {\it after} the times of maximum radial velocity.  These observed offsets are listed in the eighth column.

Why should there be any non-zero offsets?  Both the radial velocity and the brightness of the red giant are exactly tied to the orbit.  The difference is that the system has additional light from the disk that also makes a sinewave contribution that is out of phase with the ellipsoidal modulation.  This extra light comes from luminous structures in the disk, including the hot spot where the accretion column first strikes the disk.  Planquart et al. (2025) measures the positions and properties of the bright spot and the disc overflow emission, with both of these contributing light that is out of phase with the ellipsoidal modulation.  Their Figure 7 quantifies how this extra light is systematically out of phase with the orbit conjunctions.  When the variations of the hot spot are added in to the ellipsoidal modulations, the total observed brightness will be a sinewave with the peak slightly offset from the times of conjunction.  

The hot spot light varies with the accretion rate and with the band, so the offset varies with the accretion rate and the band.  This means that the observed offsets can be averaged over the measures for each state of the light curve, each with their roughly constant $\dot{M}$.  For example, in the $V$-band, I have 7 measures of the offset during the low-state, with an average of 4.48 days.  So I have adopted 4.48 days as the best offset for the $V$-band when T CrB is in the low-state.  I adopt 2.36 days as the average offset for the $V$-band in the high-state.  For the $B$-band, my adopted offsets are -0.22 days in the high-state and 3.75 days in the low state.  

For the time interval 1866--1946, the offsets are not measured.  But it is a safe assumption that the offsets for each phase are similar to those from the same phase in 1946--2025.  And this is the key to getting the pre-eruption $O-C$ curve corrected to the times of conjunction.  My adopted offsets (and their estimated uncertainties) are listed in the second from last column of Table 1.  In the end, for all the intervals with a measured $T_{\rm max}$, I have calculated the final offset $O-C$, with its propagated uncertainty, with these being tabulated in the last column of Table 1.  These final $O-C$ values are plotted in Figure 3.

Without the offsets, the observed $O-C$ curve returns a $\dot{P}$ (Schaefer 2023a) that is somewhat larger than the $\dot{P}$ from the radial velocity curve.  The offset values from 1946--2025 can only return an $O-C$ curve that follows the radial velocity $O-C$ curve by construction.  The $O-C$ curve from the ellipsoidal modulation has formal error bars that are $\sim$10$\times$ worse than for the $O-C$ curve from the radial velocity curve.  The important utility of the offset $O-C$ curve is that it covers the time interval 1866--1946.  This pre-eruption interval is required to get the critical $\Delta P$.  The pre-1946 measures are poorer than the post-1946 measures, because the 1946 eruption spurred massive observing campaigns, still on-going.  So the pre-1946 $O-C$ is scattered with large error bars.  Such is not useful for measuring $\dot{P}$ for the 1866--1946 interval.  Fortunately, even with the relatively large scatter, the final $O-C$ curve is more than adequate to measure and prove the large observed $\Delta P$.

The $\Delta P$ across the 1946 eruption will appear as a kink in 1946.  That is, there should be a parabola from 1866--1946, connecting to another parabola from 1946--2025.  This broken parabola must be continuous through 1946 (i.e., with no jumps or discontinuity in the period), because the stars will not leap forward in their orbit.  So the task is to fit a broken parabola to the $O-C$ data.  For the post-1946 interval, I have simply used the fit from the radial velocity data alone (see the previous subsection).  To be specific, the radial velocity fits for 1946--2025 give a model for $T_{\rm max}$ as
\begin{equation}
T_{\rm max} = E + N P_{\rm post} +0.5 N^2 P_{\rm post} \dot{P}_{\rm post},
\end{equation}
where $E$=2431998.282$\pm$0.64 HJD, $P_{\rm post}$=227.6043 days, and $\dot{P}_{\rm post}$=$-$3.1$\times$10$^{-6}$.  $N$ is the cycle count from the zero cycle just after the 1946 eruption and must be an integer.  The $P_{\rm post}$ value is the orbital period immediately after the 1946 eruption, for use in calculating $\Delta P$.  For 1866--1946, the fitting model is 
\begin{equation}
T_{\rm max} = E + N P_{\rm pre} +0.5 N^2 P_{\rm pre} \dot{P}_{\rm pre}.
\end{equation}
For this equation, all $N$ integers will be negative.  The $E$ value is identical to that used in Equation 4, with this ensuring no jumps in the $O-C$ curve, and it provides a solid grounding for the pre-1946 parabola.  As discussed above, $\dot{P}_{\rm pre}$ is best taken as $-$3.1$\times$10$^{-6}$, with a reasonable range from $-$1.5$\times$10$^{-6}$ to $-$4.7$\times$10$^{-6}$, and certainly $<$$+$3.1$\times$10$^{-8}$.  The fitted value of $P_{\rm pre}$ will then return $\Delta P$ as $P_{\rm post}$$-$$P_{\rm pre}$.  

Now, with the final $O-C$ measures in Table 1 and the model in Equation 5, I have performed a chi-square fit for the pre-1946 parabola.  For the most likely value of $\dot{P}_{\rm pre}$ ($-$3.1$\times$10$^{-6}$), $P_{\rm pre}$ is 227.4586, so $\Delta P$=$+$0.1457 days, or 3.5 hours.  For the 1-$\sigma$ range of $\dot{P}_{\rm pre}$, the orbital period immediately before the 1946 eruption varies from 227.4388 to 227.4770 days, for $\Delta P$ varying from $+$0.1273 to $+$0.1655 days.  For the largest possible $\dot{P}_{\rm pre}$ ($+$3.1$\times$10$^{-8}$), the pre-eruption orbital period is 227.4949, for $\Delta P$=$+$0.1211 days.  So there we have it, $\Delta P$=$+$0.146$\pm$0.019 days to appropriate precision, while $\Delta P$ is certainly $>$0.121 days.

This numerical measure of $\Delta P$ can also be seen graphically in Figure 3.  The accurate $O-C$ curve derived from the radial velocity measures is represented by the thick black curve from 1946--2025.  Closely flanking it are the curves representing the 1-$\sigma$ range of acceptable solutions, dominated by the range in $\dot{P}_{\rm post}$.  The total uncertainty in the radial velocity $O-C$ values ranges from $\pm$0.09 days in recent year, up to $\pm$0.55 days soon after the 1946 eruption.  With this, we see that the uncertainties in the post-1946 $O-C$ curve are negligibly small.  These fits have been extrapolated back in time to 1866, as displayed by the dotted black curves.  The middle dotted curve is for the extrapolation with the $\dot{P}_{\rm pre}$ equal to the best fit $\dot{P}_{\rm post}$, all with $\Delta P$=0.  These three dotted curves represent the range for the case where the sudden period change is small or zero.  Critically, all $O-C$ values from before 1910 are significantly above any reasonable extrapolation based on a zero-or-small value of $\Delta P$.  The earliest measures have the maximum brightness of T CrB occurring 20 days late, as compared to the ephemeris for $\Delta P$=0.  This is greatly larger than the uncertainties, being a 6.3-$\sigma$ deviation for the 1868 datum, and 8.6-$\sigma$ for the 1877 datum.  This shows a blatant case for the eye to see that the $\Delta P$ cannot be zero or small.  

In the $O-C$ plot, the best case for  before 1946 (i.e., with $\dot{P}_{\rm pre}$=$-$3.1$\times$10$^{-6}$) is shown by the thick black curve.  This curve is flanked by thin magenta curves representing the likely range of $\dot{P}_{\rm pre}$.  Another thin magenta curve (the lowermost in Figure 3) represents the extreme possible case (with $\dot{P}_{\rm pre}$$<$$+$3.1$\times$10$^{-8}$).  The $\Delta P$ is seen as the sharp upward kink in the curve in the year 1946.  In all possible cases, the upward kink is large and significant.  This is showing us that the T CrB $\Delta P$ in 1946 is highly significant and large and positive.

A similar result for $\Delta P$ was reported in Schaefer (2023a).  My new measure here improves on this old report in four ways:  First, the radial velocity data are used to derive a high-accuracy $O-C$ curve from 1946--2025.  This is made possible only with the excellent new radial velocity measures in Planquart et al. (2025) and Hinkle et al. (2025).  Second, this radial velocity curve is used to calibrate the offsets between times of photometric peak brightness and the maximum radial velocity, with the offsets caused by the asymmetric added light from the hotspot on the edge of the accretion disk. These offsets were applied to the same states for the 1866--1946 $O-C$ curve to get a better representation of the times of conjunction in the orbit.  Third, the radial velocity data provide a better measure of the $\dot{P}_{\rm post}$, which is then applied to $\dot{P}_{\rm pre}$.  Fourth, I apply Equations 2 and 3 to put a strong limit on $\dot{P}_{\rm pre}$, with this being critical to eliminate any possible model with a strong concave-up parabola resulting in a near-zero $\Delta P$.

This $\Delta P$ measure for T CrB is huge, with the orbital period changing by 210$\pm$27 minutes (and certainly $>$174 minutes).  This can be compared to the next highest $\Delta P$ values for any measured nova eruption, with 0.062 minutes for the 2016 eruption of U Sco (Schaefer 2023c) and 2.5 minutes for the eruption of the helium nova V445 Pup (Schaefer 2025).  That is, the sudden period change of T CrB is two orders-of-magnitude larger than for any other known nova.  If we scale the period change as $\Delta P$/$P$, then T CrB is at $+$641$\pm$83 ppm (certainly $>$531 ppm), while V445 Pup is at $+$935$\pm$27 ppm (Schaefer 2025), with the next closest nova at $+$50.3$\pm$7.9 for the last eruption of T Pyx (Schaefer 2023c).  No matter how we look at it, the T CrB $\Delta P$ is extremely large.

\section{$M_{\rm \lowercase{ejecta}}$ FOR T C\lowercase{r}B IN 1946}

Now that we have a reliable and accurate measure of $\Delta P$, we can derive $M_{\rm ejecta}$.

With Kepler's Law and the conservation of angular momentum, the period change across a nova eruption arising from the mass lost to the ejecta is
\begin{equation}
\Delta P_{\rm ml} = 2  \frac{M_{\rm ejecta}}{M_{\rm comp}+M_{\rm WD}} P
\end{equation}
(Schaefer 2020, Eq. 6).  $M_{\rm WD}$ is the WD mass and $M_{\rm comp}$ is the companion star mass.  This equation has been derived by many workers since the 1970s (it was old even for Schaefer \& Patterson 1981), and it includes the angular momentum carried away by the ejecta, which is assumed to have the same specific angular momentum as the WD.  

The observed period change ($\Delta P$) is the sum of the effects from mass loss and and the effects of `FAML',
\begin{equation}
\Delta P = \Delta P_{\rm ml} + \Delta P_{\rm FAML}.
\end{equation}
FAML is the frictional angular momentum loss caused by the companion star moving fast through the ejecta from the shell.  Since Livio, Govarie, \& Ritter (1991), FAML has been computed for the fast expanding ejecta, with this effect always being smaller in magnitude (and opposite in sign) to the mass-loss effects.  In the last few years, various workers have realized that the WD during the nova puffs up a quasi-stationary envelope from which the ejecta are launched, and the companion star plows through this envelope, with much greater effect.  Unfortunately, the density structure of this envelope is highly variable and poorly known, so detailed calculations are not now possible with any useful accuracy.  

We can know that the FAML effects should be small for various conditions in the nova system.  $\Delta P_{\rm FAML}$ is proportional to the duration of the nova eruption, i.e., scaling with the total drag of the companion star plowing through the quasi-stationary envelope surrounding the WD.  For recurrent novae, empirically, the novae are always fast or very fast, with the SyRNe having $t_3$ durations from 6 to 14 days.  $\Delta P_{\rm FAML}$ is exponentially falling with $M_{\rm WD}$ because the envelope density is governed by the scale height determined by the WD gravity.  Recurrent novae necessarily have the WD near the Chandrasekhar mass, and so their $\Delta P_{\rm FAML}$ will always be small.  $\Delta P_{\rm FAML}$ is exponentially falling with the orbital radius at which the companion star orbits the WD, because scale height of the gas in the envelope depends on the WD properties, so that a companion orbiting far from the WD will be traveling through a low density gas with little drag.  The SyRNe have very large $P$, so the red giant companion orbits far above the bulk of the envelope, leading to small drag effects.  So, for three strong reasons all pointing in the same direction, we expect the SyRNe to have small $\Delta P_{\rm FAML}$.

Critically, $\Delta P_{\rm ml}$ is always positive, and $\Delta P_{\rm FAML}$ is always negative.  So we can at least get a {\it limit} on  $\Delta P_{\rm ml}$, 
\begin{equation}
\Delta P_{\rm ml} > \Delta P.
\end{equation}
For fast long-$P$ RNe, we have only a negligibly small FAML effect, so $\Delta P_{\rm ml} \cong \Delta P$.  For slow short-$P$ novae, like for V445 Pup, the FAML effect is large (and negative), so $\Delta P_{\rm ml} \gg \Delta P$.  For novae with negative $\Delta P$, the limit is useless.  For novae with positive $\Delta P$, the limit is useful.  With Equation 6, 
\begin{equation}
M_{\rm ejecta} > 0.5 (M_{\rm comp}+M_{\rm WD})  \frac{\Delta P}{P}.
\end{equation}
This equation can usefully be applied only for two systems with giant companions, V445 Pup and T CrB.

V445 Pup is the one known example of a helium nova, and as such it is commonly suggested to be a SNIa progenitor in the SD class.  This nova had a notoriously and uniquely heavy dust cloud of ejecta, and it is a slow nova, so $\Delta P_{\rm ml} \gg \Delta P$.  V445 Pup has $M_{\rm WD}$$\gtrsim$1.35 M$_{\odot}$, while $M_{\rm comp}$ is 0.5--1.0 M$_{\odot}$.  I have discovered the orbital period to be 1.873593$\pm$0.000034 days, hence demonstrating the companion to be a giant star with surface temperature 10,000--40,000 K (Schaefer 2025).  Further, I have over a century of archival data to give $P_{\rm pre}$ and $\dot{P}_{\rm pre}$ to high accuracy, with a resultant $\Delta P$/$P$ of $+$935$\pm$27 ppm.  By Equation 9, the V445 Pup case has $M_{\rm ejecta}$$\gg$0.001 M$_{\odot}$.  The $M_{\rm accreted}$ is known from the trigger conditions to be always close to 0.00022 M$_{\odot}$ (Kato et al. 2008).  So we confidently have $M_{\rm ejecta}$$\gg$$M_{\rm accreted}$.  With this, V445 Pup has $M_{\rm WD}$ decreasing over each eruption cycle, and will never become a SNIa.  V445 Pup has been commonly held up as the exemplar of SNIa progenitors of the helium novae class, but this exemplar is now completely refuted.

T CrB is one of the fastest novae known (with $t_3$=6 days), with a very fast ejecta ($\sim$4980 km s$^{-1}$), and a very wide orbit (206 R$_{\odot}$), so the FAML must be negligibly small, so $\Delta P_{\rm ml}=\Delta P$.  T CrB has $M_{\rm WD}$=1.32$\pm$0.10 M$_{\odot}$ and $M_{\rm comp}$=0.98$\pm$0.31 M$_{\odot}$ (Planquart et al. 2025).  Along with my measured $\Delta P$/$P$ of 641$\pm$83 ppm, Equation 9 gives $M_{\rm ejecta}$=0.00074$\pm$0.00009 M$_{\odot}$. 

Is there any way to impeach this basic measurement?  An SD advocate might willfully try to impose a small-$\Delta P$, so as to get a small-$M_{\rm ejecta}$ that is smaller than $M_{\rm accreted}$.  To do so would require a near-zero $\Delta P$, visible graphically as the solution passing through 1946 with no visible kink.  Then the extension backwards in time is constrained by the physics from Equation 2 to not have any visible concave-up curvature.  So the best case for a willful SD advocate is close to the dotted black curves in Figure 3.  And these are greatly in disagreement with the observed $O-C$ measures from before 1910 (see Figures 2 and 3).  The 1868 point is 6.3-$\sigma$ above the $\Delta P$=0 dotted line, and the 1877 point is 8.6-$\sigma$ above the dotted line.  This is also quantitatively seen with the chi-square fits for the 11 pre-1946 measures, where the $\Delta P$=0 dotted curve has $\chi ^2$=145.3, versus $\chi ^2$=11.8 for the best-fit thick black curve.  So the SD advocate has no plausible means to deny the large-positive $\Delta P$ measure.  The only other possibility for an SD enthusiast is to try to impeach Equation 9.  But this is just simple physics with Kepler's Law and the requirement that $\Delta P_{\rm FAML}$$<$0.  The values for the input on the right-hand-side of Equation 9 are reasonably well measured with no chance to be 2 orders-of-magnitude smaller.  Overall, my measure of $M_{\rm ejecta}$ is just a simple timing experiment, with no dependence on the large uncertainties on extinction, filling fraction, shell volume, plus the gas temperature and density varying greatly throughout the shell.  In the end, there is no useable means to impeach or question my basic result that $M_{\rm ejecta}$ is huge.  In conclusion, we can have high confidence that $M_{\rm ejecta}$=0.00074$\pm$0.00009 M$_{\odot}$.

\section{IMPLICATIONS OF THE HUGE M$_{\rm ejecta}$}

How much mass was accreted onto the T CrB WD from 1866 to 1946?  To the needed accuracy, $M_{\rm accreted}$ can be reliably determined from two methods:  The first method is to use the luminosity of the accretion disk to measure the accretion rate as a function of year.  With this, the $M_{\rm accreted}$ has been measured to be 1.38$\times$10$^{-6}$ M$_{\odot}$ (see Table 8 of Schaefer 2023a).  

The second method is to know the mass of fresh accreted gas required to trigger the thermonuclear runaway on a WD with $M_{\rm WD}$=1.32$\pm$0.10 M$_{\odot}$ and $\dot{M}_{\rm high}$=6.4$\times$10$^{-8}$ M$_{\odot}$ yr$^{-1}$.  For the theoretical trigger mass, Yaron et al. (2005, Table 2) gives for the T CrB case that $\log [M_{\rm accreted}]$ equals $-$7.0$\pm$0.2 with units of M$_{\odot}$.  From Nomoto et al. (2007, Figure 4), for the T CrB conditions, the log-value is $-$6.4$\pm$0.3.  From Shen \& Bildsten (2009, Figure 7), for the conditions of T CrB, the log value is $-$5.5$\pm$0.2.  Shara et al. (2024) summarizes the extensive theory calculations from his group for the T CrB conditions, with the log of the trigger mass near $-$8.0 to $-$7.5.  The range of theory estimates is 2.5 orders-of-magnitude.  This range for physics calculations of the trigger mass informs us that the theory predictions of $M_{\rm ejecta}$ must have similar or larger uncertainties.  The range of theory calculations are in fine agreement with the observed log-value of $-$5.9, to within the stated error bars.  In the end, the best value is from the first method, so we have a reliable $M_{\rm accreted}$=1.38$\times$10$^{-6}$ M$_{\odot}$.

Now, we can compare $M_{\rm ejecta}$=0.00074$\pm$0.00009 M$_{\odot}$ to $M_{\rm accreted}$=1.38$\times$10$^{-6}$ M$_{\odot}$.  The ejected mass is 540$\times$ the accreted mass.  The error bars (540$\pm$65) are greatly too small to overcome this large factor.  The experimental measure is a simple timing experiment with input physics only from Kepler's Law and the impossibility of outside torques acting on the orbit.  I know of no loopholes, or exceptions, or problems, or open questions for this basic result.  So with high confidence, we have $M_{\rm ejecta}$$\gg$$M_{\rm accreted}$.

Another important implication for the large $M_{\rm ejecta}$ is that it provides a confident answer to a long-standing puzzle for nova theory, at least for the one nova T CrB.  That is, theorists have long wrestled with trying to predict the value of $M_{\rm ejecta}$, or for $M_{\rm ejecta}$/$M_{\rm accreted}$, with T CrB now providing the ground-truth for one important case.  

Historically, for decades, theorists have calculated myriads of nova-explosion models, roughly evenly divided between the WD gaining mass and the WD losing mass.  These models have huge uncertainties.  For the specific case of T CrB with a 1.35 M$_{\odot}$ ONe WD, Starrfield et al. (2024) reported five predictions of $M_{\rm ejecta}$ that range over a factor of 3300$\times$, all for simply changing the un-knowable way that the underlying WD material gets mixed in.  And their predicted $M_{\rm ejecta}$ changed by a factor of 24$\times$ simply by changing how they divided up their mass zones, from 95 zones to 300 zones.  So the results from one model from one group are extremely sensitive to free choices of unknown input, by over 4 orders-of-magnitude.  This is to say that theory has provided no useable answer.  Further, within one model for a system like T CrB, the $M_{\rm ejecta}$/$M_{\rm accreted}$ ratio flops chaotically between $<$1 and $>$1 from eruption to eruption (Hillman \& Kashi 2021).  For a comparison between competing models with a 1.35 M$_{\odot}$ ONe WD, predicted $M_{\rm ejecta}$ range from {\it zero} (!) to 0.012 to 0.42 (Starrfield et al. 2025), 3.4--4.4 (Jos\'{e} \& Hernanz 1998), and 1.9 (Rukeya et al. 2017) all in units of 10$^{-6}$ M$_{\odot}$.  All of these predictions are for old-style theory models involving only 1-dimensional calculations of the initial thermonuclear flash.  The huge range of predictions demonstrates that the old-style models have no reliability or usefulness.  

Recently, a number of independent workers have realized that all these old-style 1-D calculations are largely irrelevant anyway, because they do not include the dominant ejection mechanism.  Sparks \& Sion (2021), Shen \& Quataert (2022), and Chomiuk, Metzger, \& Shen (2020) have realized that most novae have most of their mass ejection made by the interaction of the companion star orbiting within the hot quasi-stationary envelope surrounding the WD.  This situation is complex and 3-dimensional, not yet possible to be modeled so as to get any accurate $M_{\rm ejecta}$ predictions.  There are good grounds to know that the binary-interaction with the envelope will make for large ejection masses.  In the end, neither the old-models nor the new-models can make a reliable prediction of $M_{\rm ejecta}$ that has an accuracy of better than 4 orders of magnitude.  (And recall, from Section 1 and Appendix A of Schaefer 2011, that all the traditional observational methods for measuring $M_{\rm ejecta}$ also have real error bars of 2--3 orders of magnitude.)  In this situation, the reliable and accurate measure of $M_{\rm ejecta}$ for T CrB can hopefully be used by theorists as a guide to improve their models.  

So the situation currently is that $M_{\rm ejecta}$ is not known to even within $\sim$3 orders of magnitude.  The problems are that the old-style theory had uselessly-large real error bars and completely missed the dominant ejection mechanism, the new theory that includes the dominant ejection mechanism is not yet developed enough to make any accurate prediction, and the old masses derived from hydrogen emission line fluxes have real error bars of around 2--3 orders-of-magnitude.  Unfortunately, the past expectations of our community were shaped by the olden theory and observations, without realizing the huge error bars.  

This dire situation has only one exception, and that is for the few novae with a positive measured $\Delta P$.  Currently, positive-$\Delta P$ measures are only available for T CrB, U Sco, BT Mon, V445 Pup, and T Pyx.  In general, the returned $M_{\rm ejecta}$ is only a lower limit, because we have no accurate idea as to the angular momentum loss to the binary during the eruption.  For novae with fast low-mass ejections (T CrB and U Sco), the angular momentum losses must be negligibly small, so we know the limit is really close to an equality.  For novae with long-lasting high-mass ejections (like for V445 Pup), where the angular momentum loss must be greatly larger than the mass loss effect, the $>$ sign becomes $\gg$.  Fortunately, for T CrB, we have a well-measured $\Delta P$ and an `=' sign.  That is, we have a confident and accurate measure of $M_{\rm ejecta}$ and its error bars for T CrB.  

Now, we are in a good position to replace the olden expectation as to typical $M_{\rm ejecta}$ from novae.  For the original $\Delta P$ measure on BT Mon, with the latest values from Schaefer (2023c), we have $M_{\rm ejecta}$/$M_{\rm acc}$$>$0.8.  This is just a lower limit for the only ordinary non-repeating nova in this sample.  For the unique and weird V445 Pup, we have $M_{\rm ejecta}$/$M_{\rm acc}$$\gg$5 (Schaefer 2025a).  The physical conditions for the ejection from this helium nova could be substantially different from all the other novae.  For the three recurrent novae, we have the $M_{\rm ejecta}$/$M_{\rm acc}$ ratio equal to 26 averaged over four U Sco eruptions (Schaefer 2025b), $>$8.6 for T Pyx (Schaefer 2023c), and equal to 540 for T CrB (this paper).  Both U Sco and T CrB have ONe white dwarfs, while T Pyx is not a neon nova.  These five novae span the classes of nova binaries, so large-$M_{\rm ejecta}$ appears to be general for most novae.  

With a confident $M_{\rm ejecta}$$\gg$$M_{\rm accreted}$, the T CrB WD is {\it losing mass} over each eruption cycle.  The forced implication is that the evolution of T CrB has it going from some high mass WD to a lower mass WD.  This is exactly opposite from that required by the SD scenario.  With the T CrB WD losing mass over long times, it is never going to approach the trigger condition required for a SNIa.

T CrB ejected a large mass during its 1946 eruption, so should it have produced an observable expanding nova shell?  This is a difficult question, with no practical answer.  The trouble is that expanding shells have been detected for 26 novae and these span the entire range of primary properties (Schaefer 2022).  The novae with shells are spanning the range of light curve classes (including SPODJF), spanning from FWHM velocities from 600--5350 km s$^{-1}$, spanning all spectral classes (Fe II, Hybrid, He/N and with neon), spanning decline rates with $t_3$ from 4 to 299 days, spanning orbital periods from 0.076 to 1.99 days, and spanning absolute $V$ magnitudes at peak from -3.64 to -10.38 mag.  I can find no significant correlation between a nova's properties and whether it has a discovered expanding shell.  With this, we have no indication of what makes for a detectable nova shell.  There are 6 other novae peaking at third magnitude or brighter that are shell-less, and there are 9 novae closer than T CrB that are shell-less.  Amongst bright naked eye novae, it is a mystery why DQ Her had a prominent shell, while the similar V1369 Cen had no shell.  Another conundrum is why GK Per has a famous bright shell still visible, while the similar V1494 Aql does not show any shell.  T Pyx is a recurrent nova with a prominent long-lasting shell, but no other recurrent novae have any detected shell.  Other pairs of matched novae with-and-without shells include HR Del \& V1405 Cas, CP Lac \& V630 Sgr, and V1500 Cyg \& V382 Vel.  These differences show the lack of any practical scheme for predicting whether any particular nova should or should-not have a nova shell discovered.

Nevertheless, we do have some idea as to properties that might correlate with shell visibility.  Importantly, the shell brightness should be proportional to $M_{\rm ejecta}$, so T CrB would be expected to have a bright shell on this basis.  The D- and J-class novae are expected to have ejected the most massive of nova shells (perhaps ~10$^{-4}$ M$_{\odot}$, Yaron et al. 2005), yet only 14\% have discovered shells.  So just because T CrB has a massive shell does not mean that we expect there to be a discovered shell.  The most notorious nova shells are those of the D-class, where the dust formation makes for a deep dimming in the light curve, until the expansion of the shell dilutes the extinction.  But only 17\% of D-class novae show shells.  So by this one parameter, the T CrB shell might be $\sim$10$\times$ brighter than D- and J-class shells, many of which do not make for detected shells.  The brightness and density in the shell should scale inversely as the duration of the ejection, which is comparable to $t_3$.  The massive-shelled D- and J-class novae have median $t_3$ of 83 days, while T CrB has a $t_3$ of 6 days. So with this scaling, with all else being equal, T CrB should have a 14$\times$ fainter shell than the D- and J-class novae.  The shell density will scale as the inverse-cube of the expansion velocity, and the expansion velocity scales as the FWHM of the H$\alpha$ line.  Amongst D-class and J-class novae with massive shells, the FWHM ranges between 300 and 3599 km s$^{-1}$ (with a median of 1050 km s$^{-1}$).  T CrB has a FWHM of 4980 km s$^{-1}$.  So the shell brightness for T CrB should be around 100$\times$ dimmer than for the other massive shells, on this basis alone.  Collecting these scalings, the brightness of the T CrB shell, as compared to the D- and J-class novae, will be roughly 10$\times$ brighter due to the ejecta mass, 14$\times$ fainter due to the eruption duration, and 100$\times$ fainter due to the expansion velocity.  With these simplistic scalings, the T CrB shell would be greatly fainter than for D- and J-class novae, most of which were invisible in practice.  So it is reasonable to expect that the T CrB shell would not be visible.

An implication of the huge $M_{\rm ejecta}$ and the long-term decrease in $M_{\rm WD}$ is that the T CrB WD must be an ONe WD.  The reason is that the original formation of the WD must have had the WD more massive than 1.32$\pm$0.10 M$_{\odot}$, and such is only possible for a ONe WD.  That is, CO WDs can only form with original masses $<$1.2 M$_{\odot}$, so the observed decrease in mass cannot produce the observed T CrB WD.  ONe WDs can only form with original $M_{\rm WD}$$>$1.2 M$_{\odot}$.  So the scenario is that the T CrB WD formed at 1.33 or 1.40 M$_{\odot}$, and has now whittled its mass down to 1.32 M$_{\odot}$ or so.  This demonstrates that the T CrB WD can only be of ONe composition, and such WDs cannot become a Type Ia supernova.

What is the scenario for T CrB evolution as a recurrent nova?  T CrB started out as a wide binary with a $\sim$8 M$_{\odot}$ star plus a 1--2 M$_{\odot}$ star, the more massive star went through a planetary nebula stage that tightened the orbit, leaving behind a near-maximal ONe WD.  After some time, the secondary star starts its ordinary expansion to a red giant.  Relatively recently, the red giant fills its Roche lobe, and accretion starts, accelerating over time.  At some point, gas accumulated on the surface of the WD reaches the trigger threshold, and T CrB erupts as a classic nova.  Over time, the $\dot{M}$ increases and the recurrence time scale decreases.  Perhaps this increasing $\dot{M}$ is the cause of the apparent decrease in the recurrence timescale from 1217 to this year?  It is still unknown what is driving the accretion rate, because the ordinary expansion of the red giant cannot drive such a high accretion rate as is now seen, and we do not know of any angular momentum loss mechanism that can shrink the Roche lobe size at a fast enough rate.  I am thinking that this mystery is fundamental for T CrB, and that we will not really understand T CrB until this mystery is solved.

T CrB has been erupting since at least the year 1217, with 10 eruptions from 1217 to 1946.  How many years has T CrB been erupting as a nova?  Over the last hundred or thousand eruptions, the T CrB ejecta has been plowing into the Interstellar Medium and building up a `Nova Super-remnant' (Shara et al. 2024).  T CrB appears near the poorly-defined center of the super-remnant.  The {\it Gaia} proper motion has T CrB traveling along a radius of the super-remnant in $\sim$200,000 years.  The age of this super-remnant can constrain the number of years since T CrB started its nova eruptions.  If T CrB started erupting 200,000 years ago, when it was situated on the edge of the super-remnant, then there is no chance that all the plowed-up ISM would make a super-remnant that is centered near the current position of T CrB.  So the super-remnant must be greatly {\it younger} than 200,000 years.  As shown in Figures 3 and 4 of Shara et al. (2024), T CrB is within $\sim$50,000 years of proper motion of the ill-defined center of the super-remnant.  This means that we can only place an upper limit on the age, and that limit is $\sim$50,000 years.  So the observed proper motion and the position of the super-remnant only restricts the duration of eruptions to be $<$50,000 years or so.

How has the recurrence timescale ($\tau_{\rm rec}$) been changing?  Well, the time span from 1217.8 to 1787.9 is 7$\times$81.4 years.  The recurrence time from 1787.9 to 1946.1 is 2$\times$79.1 years.  Apparently, $\tau_{\rm rec}$ has been speeding up over the last 8 centuries.  Fitting a linear trend to the observed eruption dates gives a change of $-$0.45 years for every eruption.  With an intrinsic scatter of 0.4 years, the best-fit linear trend has $\chi^2$=3.6, while the best-fit case with no change in $\tau_{\rm rec}$ has $\chi^2$=115.3.  So the change in $\tau_{\rm rec}$ is highly significant.  Given the observed eruption dates, the best estimate is that the eruption cycle shortens by 0.45 years each cycle.

Given this trend in $\tau_{\rm rec}$, if extrapolated back by 100 eruptions, the recurrence time was 125 years.  The average over the last 100 eruptions would be just over 100 years between eruptions, and these 100 eruptions would have taken nearly 10,000 years.

How many times has T CrB been erupting as a nova?  For this, the tightest constraint comes from the limitation on how much mass has been lost by the WD.  The original ONe WD must have started out with a mass $<$1.40 M$_{\odot}$ or so.  The current mass of the WD is 1.32$\pm$0.10 M$_{\odot}$, and the WD is losing mass at the rate of 0.00074$\pm$0.00009 M$_{\odot}$ each eruption.  Taking the best values, the maximum number of eruptions is (1.40-1.32)/0.00074 = 108.  With the error bars, the number of eruptions is $<$108$\pm$135, while a further observed limit from the 1217 eruption requires that the number is also $>$10.

So we now have a simple and inevitable picture of the past evolution of the T CrB binary.  The original $\sim$8 M$_{\odot}$ star evolved into an ONe WD close to the Chandrasekhar mass.  After the companion star expanded as a red giant to fill its Roche lobe, the accretion started, and the classical nova eruptions started.  The increasing accretion makes for a shortening $\tau_{\rm rec}$, which has `recently' turned T CrB into a recurrent nova with eruptions coming faster than once-per-century.  For the last 8 centuries, we are seeing T CrB shortening the eruption cycle by 0.45 years each cycle.  The WD mass is being whittled down, arriving at its current mass with something like 100 eruptions.  With this, T CrB started it nova events roughly 10,000 years ago.  That the T CrB $M_{\rm WD}$ is decreasing over time and that the WD has an ONe composition, T CrB certainly cannot be a SNIa progenitor.

\section{T C\lowercase{r}B AND THE SNI\lowercase{a} PROGENITOR PROBLEM}

T CrB certainly has a large and positive $\Delta P$, with $\Delta P$/$P$ equal to $+$641$\pm$83 ppm (and certainly $>$531 ppm).  With simple and certain physics, this $\Delta P$ is used to calculate that the ejected mass during the 1946 nova eruption was huge, with $M_{\rm ejecta}$=0.00074$\pm$0.00009 M$_{\odot}$.  This can then be compared to the reliable and accurate measure that the T CrB white dwarf accreted $M_{\rm accreted}$=1.38$\times$10$^{-6}$ M$_{\odot}$ from 1866 to 1946.  We then have $M_{\rm ejecta}$$\gg$$M_{\rm accreted}$ by a factor of 540$\times$.  With this, the T CrB WD in 1946 ejected greatly more mass than it accreted over the previous eruption cycle, so necessarily $M_{\rm WD}$ is getting smaller fast.  With $M_{\rm WD}$ {\it decreasing} over each eruption cycle, the WD can never approach any condition that would trigger a Type Ia supernova.  That is, T CrB is certainly {\it not} a SNIa progenitor.

Further, T CrB has an ONe WD.  We know this for two independent reasons.  The first is that the neon-to-oxygen emission line flux ratio is $F_{3869}/F_{5007}$=1.4.  The flux ratio $\ge$0.3 is a perfect indicator for neon novae, with such requiring an ONe WD.  The second reason is that the T CrB WD must have started out with more mass than 1.32$\pm$0.10 M$_{\odot}$, and this is only possible for an ONe WD.  With it being impossible for an ONe WD to explode as a SNIa, we know that T CrB is not a SNIa progenitor.

T CrB has long been the prototype SD progenitor, either as a SS or RN system.  Now, I have a strong proof that T CrB cannot possibly become a SNIa.  This is a severe blow for single-degenerate models as solutions to the SNIa progenitor problem.

\begin{acknowledgments}
I thank Joanna Miko{\l}ajewska (Nicolaus Copernicus Astronomical Centre), L\'{e}a Planquart (Institut d'Astronomie et d'Astrophysique, Universit\'{e} Libre de Bruxelles), and Kenneth Hinkle (National Optical-Infrared Astronomy Research Laboratory) for comments and help on the science content. 
\end{acknowledgments}


%\vspace{5mm}
%\facilities{ }


{}

\end{document}